\documentclass[12pt]{article}
\usepackage{epsfig,euscript,latexsym,amsfonts,amsmath,amsthm,amssymb}

\title{Quantum-like Representation Algorithm: Transformation of Probabilistic Data into vectors on Bloch's  Sphere}

\author{Andrei Khrennikov\\
International Center for
Mathematical Modeling \\ in Physics and Cognitive Sciences\\
University of V\"axj\"o, S-35195, Sweden}

\begin{document}
\maketitle

\abstract{In this paper we present a simple algorithm  for
representation of statistical data of any origin by complex
probability amplitudes. Numerical simulation with Mathematica-6 is
performed.  The Bloch's sphere is used for visualization of
results of numerical simulation. On the one hand, creation of such
a quantum-like (QL) representation and its numerical approval is
an important step in clarification of extremely complicated
interrelation between classical and quantum randomness. On the
other hand, it opens new possibilities for application the
mathematical formalism of QM in other domains of science.}



\section{Introduction}

The problem of interrelation between classical and quantum
randomness is extremely complicated, see von Neumann \cite{VN} for
the pioneer study. It has not yet been solved completely, see e.g.
Svozil \cite{Svozil}, \cite{Svozil1} for discussions.  By the
orthodox Copenhagen viewpoint these are two completely different
types of randomness. The gap between them is huge. However, we can
mention a number of attempts to minimize this gap, e.g.
introduction of Wigner distribution and later development of
quantum tomography, see e.g. \cite{MM}, \cite{MD} and references
hereby, see also \cite{Raedt} on recent numerical simulation for
quantum correlations. Mentioned studies were oriented toward
classical probabilistic interpretation of quantum randomness.

In this paper we would like to formulate and study "inverse Born's
rule problem" -- {\bf IBP}: to construct a representation of
probabilistic data by probability amplitudes which match Born's
rule. In \cite{KH}, \cite{KH1} we proposed an algorithm --
quantum-like representation algorithm (QLRA) -- which transforms
probabilistic data of any origin in probability amplitudes
(complex and hyperbolic). In this paper we simplify QLRA
essentially (in particular, by restricting its domain of
application). We performed extended numerical simulation based on
Mathematica-6. The Bloch's sphere is used for visualization of
results of numerical simulation. On the one hand, creation QLRA
and its numerical approval is an important step clarification of
extremely complicated interrelation between classical and quantum
randomness. On the other hand, it opens new possibilities for
application the mathematical formalism of QM in other domains of
science, see e.g. \cite{H1}, \cite{AB}. By representing
probabilistic data by QL-states one might try to apply methods of
quantum information theory to study such data. Of course, only the
first step has been done in this direction.

\section{Inversion of Born's rule}

We consider the simplest situation. There are given two
dichotomous observables $a=\alpha_1, \alpha_2$ and $b=\beta_1,
\beta_2.$ They can be physical (classical or quantum) observables
or e.g. two questions which are used for tests in psychology,
cognitive or social science and so on. We set $X_a=\{\alpha_1,
\alpha_2\}$ and $X_b=\{\beta_1, \beta_2\}$ -- "spectra of
observables."

We suppose that there is given  the matrix of transition
probabilities ${\bf P}^{b\vert a}= (p_{\beta \alpha}),$ where
$p_{\beta \alpha}\equiv P(b=\beta \vert a=\alpha)$ is the
probability to obtain the result $b=\beta$ under the condition
that the result $a=\alpha$ has been obtained.

There are also given probabilities $p^a_\alpha \equiv P(a=\alpha),
\alpha\in X_a,$ and $p^b_\beta \equiv P(b=\beta), \beta \in X_b.$
Probabilistic  data $C=\{ p^a_{\alpha}, p^b_{\beta}\}$ is related
to some experimental context (in physics preparation procedure).
It is not assumed that both observables can be measured
simultaneously. Thus in general two samples (prepared under the
same complex of conditions) are used to collect the probabilistic
data for observations of  $a$ and $b,$ respectively.

Our aim is to represent this data by a probability amplitude
$\psi)$ (in the simplest case it is complex valued) such that
Born's rule  holds for both observables:
\begin{equation}
\label{BR} p^b_{\beta}= \vert \langle \psi, e_\beta^b\rangle
\vert^2,\; p^a_{\alpha}= \vert \langle \psi, e_\alpha^a\rangle
\vert^2,
\end{equation}
where $\{e_\beta^b\}_{\beta \in X_b}$ and $\{e_\alpha^a\}_{\alpha
\in X_a}$ are orthonormal bases (which are also produced by QLRA)
for observables $b$ and $a,$ respectively. These observables are
represented by operators $\widehat b$ and $\widehat a$ which are
diagonal in these bases.

\section{QLRA}

In \cite{KH}, \cite{KH1} we derived the following formula for
interference of probabilities:
\begin{equation}
\label{TFR} p^b_{\beta} = \sum_\alpha p^a_{\alpha} p_{\beta
\alpha} + 2 \lambda_\beta \; \sqrt{\prod_\alpha p^a_{\alpha}
p_{\beta \alpha}},
\end{equation}
where the "coefficient of interference"
\begin{equation}
\label{KOL6} \lambda_\beta = \frac{p^b_{\beta} - \sum_\alpha
p^a_{\alpha} p_{\beta \alpha}}{2 \sqrt{\prod_\alpha p^a_{\alpha}
p_{\beta \alpha}}} .
\end{equation}
This is a trivial mathematical identity. To prove it, one should
just put $\lambda$ given by (\ref{KOL6}) into (\ref{TFR}). A
similar representation we have for the $a$-probabilities. To
simplify considerations, we shall proceed under the condition:

{\bf DS}: The matrix of transition probabilities ${\bf P}^{b \vert
a}$ is doubly stochastic.\footnote{In a doubly stochastic matrix
all entries are nonnegative and all rows and all columns sum to
1.}

We also suppose that probabilistic data $C=\{ p^a_{\alpha},
p^b_{\beta}\}$ consists of strictly positive probabilities. We
proceed under the basic assumption (specifying the type of
representation):

{\bf RC}: Coefficients of interference $\lambda_\beta, \beta \in
X_b,$ are bounded by one:
$$
\vert \lambda_\beta\vert \leq 1.
$$
Probabilistic data $C$ such that {\bf RC}  holds is called {\it
trigonometric}, because in this case we have the conventional
formula of trigonometric interference:
\begin{equation}
\label{TNCZ} p^b_{\beta} = \sum_\alpha p^a_{\alpha} p_{\beta
\alpha} + 2 \cos\phi_\beta \; \sqrt{\prod_\alpha p^a_{\alpha}
p_{\beta \alpha}},
\end{equation}
where
$$
\lambda_\beta=\cos \phi_\beta.
$$
This is simply a new parametrization: a new parameter $\phi$ is
used, instead of $\lambda.$  Parameters $\phi_\beta$ are said to
be $b \vert a$-{\it relative phases} for the data $C.$ We defined
these phases purely on the basis of probabilities.\footnote{We
have not started with any linear space; in contrast we shall
define geometry from probability. In the conventional quantum
formalism the formula of interference of probabilities is derived
starting directly with the Hilbert space. We recall that in QM
interference of probabilities is derived via transition from the
basis for the $a$-observable to the basis for the $b$-observble.
From the very beginning observables are given by self-adjoint
operators.}

From the probabilistic viewpoint the formula for interference of
probabilities is nothing else than perturbation of the classical
formula of total probability (FTP). We recall this law in the
simplest case of dichotomous random variables
\begin{equation}
\label{F} P(b=\beta)= P(a=\alpha_1) P(b= \beta \vert a=\alpha_1) +
P(a=\alpha_2) P(b= \beta \vert a=\alpha_2)
\end{equation}
Thus the probability $P(b=\beta)$ can be reconstructed on the
basis of conditional probabilities $P(b=\beta \vert a=\alpha).$
FTP plays the fundamental role in classical statistics and
decision making. However, it is violated in experiments with
quantum systems.

We denote the collection of  trigonometric probabilistic data by
the symbol ${\cal C}^{\rm{tr}}.$ By using the elementary formula:
$ D=A+B+2\sqrt{AB}\cos \theta=\vert \sqrt{A}+e^{i
\phi}\sqrt{B}|^2, $ for real numbers $A, B > 0, \theta\in [0,2
\pi],$ we can represent the probability $p^b_{\beta}$ as the
square of the complex amplitude (Born's rule):
\begin{equation}
\label{Born} p^b_{\beta}=\vert \psi(\beta) \vert^2 \;.
\end{equation}
Here
\begin{equation}
\label{EX1} \psi(\beta) = \sqrt{p^a_{\alpha_1}p_{\beta \alpha_1}}
+ e^{i \phi_\beta} \sqrt{p^a_{\alpha_2} p_{\beta \alpha_2}}, \;
\beta \in X_b.
\end{equation}

The formula (\ref{EX1}) gives the quantum-like representation
algorithm -- QLRA. For any trigonometric probabilistic data $C$
QLRA produces the complex amplitude $ \psi.$ This algorithm can be
used in any domain of science to create the QL-representation of
probabilistic data.

We denote the space of functions: $\psi: X_b\to {\bf C}$ by the
symbol $\Phi =\Phi(X_b, {\bf C}).$ Since $X= \{\beta_1, \beta_2
\},$ the $\Phi$ is the two dimensional complex linear space. By
using QLRA
 we construct the map $J^{b \vert a}:{\cal C}^{\rm{tr}}
\to \Phi(X, {\bf C})$ which maps probabilistic data into complex
amplitudes. The representation ({\ref{Born}}) of probability is
nothing else than the famous {\it Born rule.} The complex
amplitude $\psi(\beta)$ can be called a {\it wave function} of the
data $C$ or a  (pure) {\it state.}

By using the terminology of quantum information theory we can say
that QLRA represents probabilistic data (of a special sort,
namely, trigonometric) by {\it qubits.}

We set $e_\beta^b(\cdot)=\delta(\beta- \cdot)$ -- Dirac
delta-functions concentrated in points $\beta= \beta_1, \beta_2.$
The Born's rule for complex amplitudes (\ref{Born}) can be
rewritten in the following form: $$\label{BH} p^b_{\beta}=\vert
\langle \psi, e_\beta^b \rangle \vert^2,$$ where the scalar
product in the space $\Phi(X_b, C)$ is defined by the standard
formula:
\begin{equation}
\label{SPR} \langle \psi_1, \psi_2 \rangle = \sum_{\beta\in X_b}
\psi_1(\beta)\bar \psi_2(\beta).
\end{equation}
 The system of functions
$\{e_\beta^b\}_{\beta\in X_b}$ is an orthonormal basis in the
Hilbert space $H=(\Phi, \langle \cdot, \cdot \rangle).$

Let now $X_b \subset {\bf R}$ (in general $\beta$ is just a label
for a result of observation). By using the Hilbert space
representation  of the Born's rule we obtain the Hilbert space
representation of the expectation of the observable $b: E b =
\sum_{\beta\in X_b} \beta\vert\psi(\beta)\vert^2= \sum_{\beta\in
X_b} \beta \langle \psi, e_\beta^b\rangle \overline{\langle\psi,
e_\beta^b\rangle}= \langle \hat b \psi, \psi\rangle,$
 where the
(self-adjoint) operator $\widehat b: H \to H$ is determined by its
eigenvectors: $\widehat b e_\beta^b=\beta e^b_\beta, \beta\in
X_b.$ This is the multiplication operator in the space of complex
functions $\Phi(X_b,{\bf C}):$ $ \widehat{b} \psi(\beta) = \beta
\psi(\beta).$

To solve IBP completely, we would like to have Born's rule not
only for the $b$-variable, but also for the $a$-variable:
$p^a_{\alpha}=\vert \langle \psi, e_\alpha^a \rangle\vert^2 \;,
\alpha \in  X_a.$ How can we define the basis $\{e_\alpha^a\}$
corresponding to the $a$-observable? Such a basis can be found
starting with interference of probabilities.  We have:
\begin{equation} \label{0} \psi=\sqrt{p^a_{\alpha_1}}
f_{\alpha_1}^a + \sqrt{p^a_{\alpha_2}} f_{\alpha_2}^a,
\end{equation}
where
\begin{equation}
\label{Bas} f_{\alpha_1}^a =\left( \begin{array}{l} \sqrt{p_{\beta_1 \alpha_1}} \\
\sqrt{p_{\beta_2 \alpha_1}}
\end{array}
\right ),\; \; f_{\alpha_2}^a =\left( \begin{array}{l} e^{i\phi_{\beta_1}} \sqrt{p_{\beta_1 \alpha_2}} \\
e^{i \phi_{\beta_2}} \sqrt{p_{\beta_2 \alpha_2}}
\end{array}
\right )
\end{equation}
The condition {\bf DS}  implies that the  system of vectors
$\{f_{\alpha_i}^a\}$ is an orthonormal basis iff the probabilistic
phases satisfy the constraint:
$$
\phi_{\beta_2} - \phi_{\beta_1}= \pi \; \rm{mod} \; 2 \pi,
$$
Thus, instead of the $a$-basis (\ref{Bas}) which depends on
phases, we can consider a new $a$-basis which depends only on the
matrix of transition probabilities:
\begin{equation}
\label{BasT} e_{\alpha_1}^a =\left( \begin{array}{l} \sqrt{p_{\beta_1 \alpha_1}} \\
\sqrt{p_{\beta_2 \alpha_1}}
\end{array}
\right ),\; \; e_{\alpha_2}^a =\left( \begin{array}{l}  \; \sqrt{p_{\beta_1 \alpha_2}} \\
- \sqrt{p_{\beta_2 \alpha_2}}
\end{array}
\right )
\end{equation}
The $a$-observable is represented by the operator $\hat{a}$ which
is diagonal with eigenvalues $\alpha_1,\alpha_2$ in the basis
$\{e_{\alpha}^a\}.$

\section{Numerical simulation and visualization on Bloch's sphere}

\begin{figure}[ht]
\begin{center}
\epsfig{file=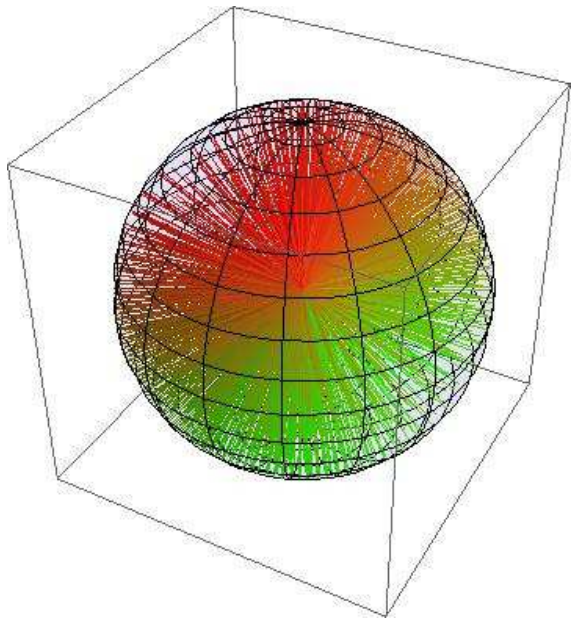,width=7.5cm} \caption{Transition probability
$P=0.1$} \label{fig:P03}
\end{center}
\end{figure}

We recall that at the moment we work under condition {\bf RC}.
First of all our computer program checks this condition. If {\bf
RC} is violated, then the program's output is empty -- no point on
Bloch's sphere. We use spherical coordinates: $x=\sin 2\theta \cos
\phi, y= \sin 2\theta \sin \phi, z= \cos 2\theta.$ A vector on
Bloch's sphere is given by
\begin{equation}
\label{BSV}
\psi= \cos \theta \vert 0\rangle + \sin \theta e^{i \phi} \vert 1\rangle
\end{equation}
It is convenient  to use the QLRA-output  in the $a$-basis. Thus
we make identification: $\vert 0\rangle=e^a_{\alpha_1} ,\; \vert
1\rangle =e^a_{\alpha_2}.$ We have: $p^a_{\alpha_1}=\cos\theta, \;
p^a_{\alpha_2}=\sin \theta; \lambda_{\beta_1}= \cos \phi, \; \pm
\sqrt{1- \lambda_{\beta_1}^2}= \sin \phi.$ Finally:
\[x=2 \sqrt{p^a_{\alpha_1}p^a_{\alpha_2}} \lambda_{\beta_1}\]
\[y=\pm 2 \sqrt{p^a_{\alpha_1}p^a_{\alpha_2}} \sqrt{1-\lambda_{\beta_1}^2}\]
\[z= p^a_{\alpha_1}- p^a_{\alpha_2}\]

\begin{figure}[ht]
\begin{center}
\epsfig{file=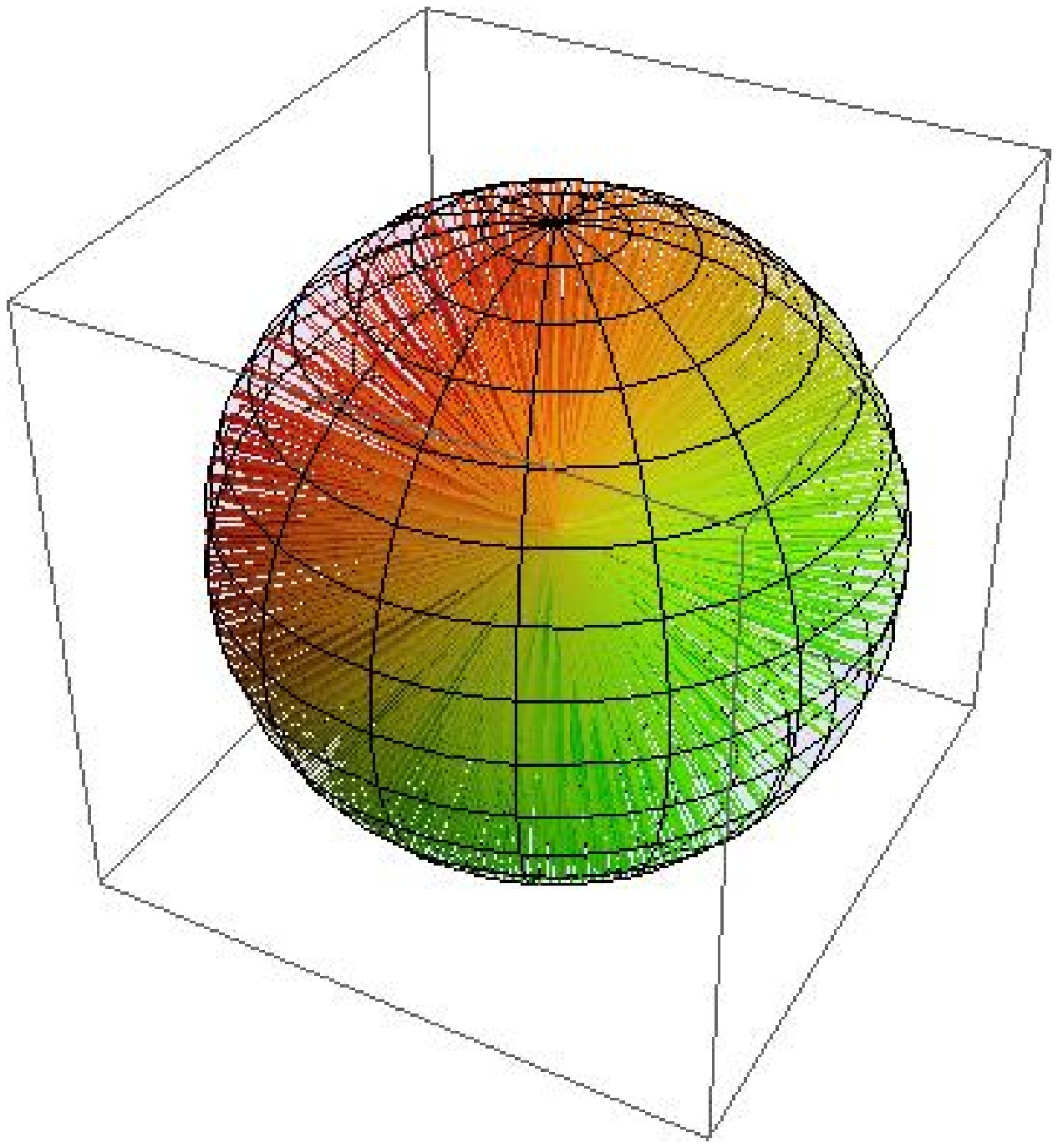,width=7.5cm} \caption{Transition probability $P=0.5$}
\label{fig:P05}
\end{center}
\end{figure}

In the program\footnote{Simulation was done with help (in
programming and visualization) of I. Basieva during her visit to
International Center for Mathematical Modelling of University of
V\"axj\"o in December 2007.} we  make the parametrization of
probabilities: $p^a_{\alpha_1}= q, p^a_{\alpha_2}=1-q,
p^b_{\beta_1}=p, p^b_{\beta_1}=1-p.$ Bigger values of $q$ give
more red color, larger values of $p$ give more green. If both $q$
and $p$ are rather small, then picture is dark. If both $q$ and
$p$ are rather big, then picture is yellow. Since we proceed under
condition {\bf DS}, the elements of the matrix of transition
probabilities can be parameterized as $p_{\beta_1
\alpha_1}=p_{\beta_2 \alpha_2}= P, \; p_{\beta_1 \alpha_2}=
p_{\beta_2 \alpha_1}= 1-P.$

\end{document}